\documentclass[aps,prl,twocolumn,groupedaddress,showpacs]{revtex4-1}

\usepackage{color}
\usepackage{amsmath}
\usepackage{graphicx}
\usepackage{subfigure}

\begin{document}

\title{Update rules and interevent time distributions: Slow ordering vs. no ordering in the Voter Model}

\author{J. Fern\'andez-Gracia}
\email[]{juanf@ifisc.uib-csic.es}
\author{V.M. Egu\'iluz}
\author{M. San Miguel}
\affiliation{ IFISC, Instituto de F\'isica interdisciplinar i Sistemas Complejos (CSIC-UIB), Campus Universitat Illes Balears, E-07122 Palma de Mallorca, Spain}

\date{\today}

\begin{abstract}
We introduce a general methodology of update rules accounting for arbitrary interevent time distributions in simulations of interacting agents. In particular we consider update rules that depend on the state of the agent, so that the update becomes part of the dynamical model.  As an illustration we consider the voter model in fully-connected, random and scale free networks with an update probability inversely proportional to the persistence, that is, the time since the last event. We find that in the thermodynamic limit, at variance with standard updates, the system orders slowly. The approach to the absorbing state is characterized by a power law decay of the density of interfaces, observing that the mean time to reach the absorbing state might be not well defined.
\end{abstract}


\maketitle

{\em Introduction.-} There exists broad empirical evidence of a large heterogeneity in the timing of individual activities \cite{malmgrem_univ_corr, Dar_Eins_BAr, origin_bursts, model_bursts, moro,  small_but_slow, malmgrem_poiss, Eckmann_orig,model_scaling}: The distribution of times at which an individual initiates an action or an interaction with another individual is such that a meaningful characteristic interevent time does not exist. Beyond studying the origin of such individual interevent distributions \cite{Dar_Eins_BAr,vazquez_non_poiss,origin_bursts,model_bursts}, an important challenge is to study the implications of these activity patterns in the collective behavior of interacting agents. Much understanding of collective behavior has been gained by simulation of models of interacting agents. The standard approach in these simulations is that agents are updated following independent random Poisson processes (random asynchronous update) so that an exponential interevent time distribution is expected. The characteristic updating time is the Monte Carlo step in which each agent has been updated once on average. It is known that other updating mechanisms which also rely on a characteristic time, such as the extreme case of synchronous update, can result in different collective behavior of the system or even artificial behavior \cite{Nowak,Hubermann}. Still, the necessary implementation in simulation studies of updating mechanisms that incorporate the observed heterogeneous timing of agents actions is largely unexplored. This is the central issue addressed in this paper.

The effects of power law or broad interevent distributions in the collective behavior have been, so far, mostly addressed in the context of spreading of information or infection processes \cite{moro,vazquez_non_poiss,small_but_slow,vazquez_spread} resulting in a slowing down of the dynamics. Here we will focus on consensus processes \cite{opinion,castellano_stat_phys} in which agents can be in several equivalent states, and instead of a spreading process, there is a competition between these states. The interaction among agents leads either to a consensus in one of these states or to asymptotic coexistence of different states. The system is said to order when in the thermodynamic limit there is unbounded growth of the number of agents in one of the equivalent states. In dealing with these consensus processes we propose a general updating algorithm and implement it in two conceptually different ways. Updating means here the attempt to change the state of the agent according to an interaction with her neighbors, so that updating does not necessarily mean change of state \cite{tessone,Takaguchi10}. In both implementations considered, the update probability depends on an internal time, giving rise to a heterogeneous timing. In a first case, \textit{exogenous update}, the update probability is independent of the state of the agent but in a second case, \textit{endogenous update}, it is a function of the persistence time of that agent, \textit{i.e.}, the time spent since its last change of state. We propose this second updating mechanism as a genuine ingredient for the understanding of many aspects of social collective behavior: There is a co-evolution of the state of the agent and the updating algorithm, so that the updating process is itself a part of the dynamical model of agent-agent interaction. We will argue that qualitative changes in the collective behavior, as for example ordering vs non-ordering in consensus processes, occur due to such state-dependent update. As an illustration of these general problems we will consider the simplest consensus problem described by the voter model \cite{holley_liggett,opinion,castellano_stat_phys}.

{\em The update.-} A set of $N$ agents are placed on the nodes of a network of interaction. Each agent $i$ is characterized by its state $s_i$ and an internal variable that we will call \textit{persistence time} $\tau_i$. For any given interaction model (Ising, voter, contact process, ...), the dynamics is as follows: at each time step,
\begin{enumerate}
 \item with probability $p(\tau_i)$ each agent $i$ becomes active, otherwise it stays inactive;
 \item active agents update their state according to the dynamical rules of the particular interaction model;
 \item all agents increase their persistence time $\tau_i$ in one unit
\end{enumerate}

\begin{figure}[ht]
\includegraphics[height=0.48\textwidth,angle=-90]{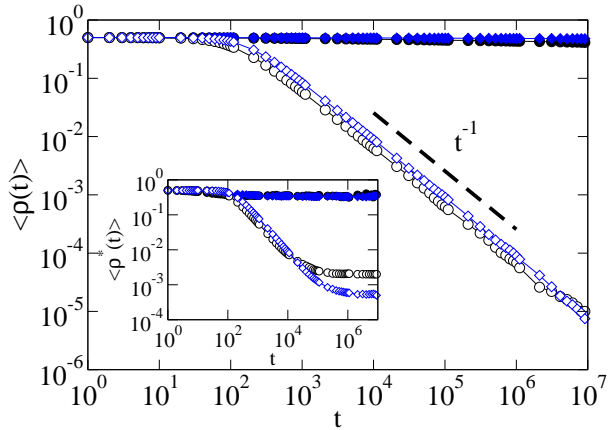}
  \caption{\label{nocoev_update_rho_meanf}Average density of interfaces $\langle\rho(t)\rangle$ with activation probability $p(\tau)=1/\tau$ on a fully connected network [Inset: density of interfaces averaged over surviving runs $\langle\rho^*(t)\rangle$]. Empty symbols stand for the endogenous update while filled ones for the exogenous. System sizes are 1000 (circles) and 4000 (diamonds); average over 1000 realizations.}
\end{figure}

The persistence time measures the time since the last event for each agent. Typically an event is an interaction (\textit{exogenous update}: active agents reset $\tau =0$ after step (ii)) or a change of state (\textit{endogenous update}: only active agents that change their state in step (ii) reset $\tau =0$).

There are two interesting limiting cases of this update when $p(\tau)$ is independent of $\tau$: when $p(\tau)=1$, all agents are updated synchronously; when $p(\tau)=1/N$, every agent will be updated on average once per $N$ unit time steps. The latter corresponds to the usual random asynchronous update (RAU). We are interested in non-Poissonian activation processes, with probabilities $p(\tau)$ that decay with $\tau$, that is, the longer an agent stays innactive, the harder is to activate. To be precise, we will later consider that
\begin{equation}
p(\tau) = \frac{b}{\tau}~,\label{pdetau}
\end{equation}
where $b$ is a parameter that controls the decay with $\tau$.

For the sake of clarity we will focus on the voter model \cite{holley_liggett,opinion,castellano_stat_phys}. In this model the state of an agent can take only one of two values $+1$ or $-1$. Initially the state of the agents is randomly assigned any of the two possibilities and the internal persistence time $\tau_i$ is set to $0$ for all agents. At step (ii) above, each active agent updates its state copying the state of one of its neighbors chosen at random $s_i(t) = s_j(t-1)$ (node update dynamics).

The voter model has two absorbing configurations corresponding to all agents having the same state. The absorbing state is reached in a finite time in any finite network (as long as all nodes are reachable from at least one node). The approach to the absorbing state can be characterized by the time evolution of the ensemble average of the density of interfaces $\langle\rho(t)\rangle$, that is, links in the network connecting nodes with different states. The voter model has been usually studied with RAU dynamics. A main conclusion in these studies is that the qualitative form of the evolution of $\langle\rho(t)\rangle$ depends on the effective dimensionality $d$ of the interaction network \cite{SucheckiPRE}. For $d\le 2$ the system orders, so that there is a coarsening process with growth of domains of agents in the same state. However, for $d>2$ the RAU voter dynamics does not order the system, and there is no coarsening process. In each dynamical realization $\rho(t)$ reaches a plateau value associated with a long lived dynamically active state until a finite size fluctuation takes the system to the absorbing state. This is reflected in an exponential decay for the ensemble average $\langle\rho(t)\rangle$. The characteristic time of the exponential decay is found to scale linearly with system size $\tau \sim N$ for a fully connected network and for general random uncorrelated networks \cite{vazquez_eguiluz}, while for a Barab\'asi-Albert scale-free network $\tau \sim N\ln N$ \cite{conserv_sucheki,voter_redner,Castellano05}. We have performed detailed numerical simulations checking that the same exponential decay for $\langle\rho(t)\rangle$ and the same system size scaling of the characteristic times is found in these networks when using a synchronous update or a sequential asynchronous update instead of a RAU.

\begin{figure}
 \includegraphics[height=0.48\textwidth,angle=-90]{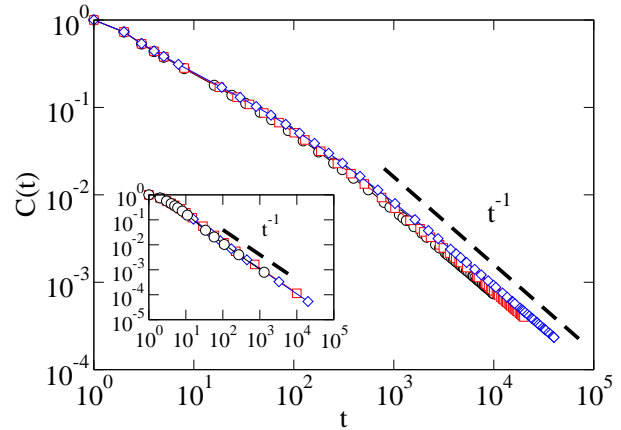}
 \caption{Cumulative persistence distribution for the endogenous update [inset: for the exogenous update] with activation probability $p(\tau)=1/\tau$ on a fully connected network of sizes 1000 (circles), 2000 (squares) and 4000 (diamonds).}
 \label{cdetau_meanf_coup_uncoup}
\end{figure}

The concept of persistence has been quantified by measuring the number of agents that have not changed their state at time $t$ starting from a given initial condition. This has been analyzed in detail in the voter model in regular lattices and in a mean-field approach \cite{ben_naim_coarsening_persistence}. This quantity is found to be a good characterization of the dynamics, showing for example different behavior for voter and Ising dynamics. For the mean field solution of the voter model it is found that the fraction of agents that have not changed their state after $t$ time steps decays exponentially with a characteristic time that depends on the initial fraction of agents in each state. In order to compare with recent empirical analysis, we extend this concept of persistence and measure the probability that an agent changes its state after $t$ time steps, $M(t)$. Equivalently the cumulative distribution  $C(t)=1-\int_1^{t}M(T)dT$ measures the probability that an agent has not changed its state after $t$ time steps. We have performed detailed numerical simulation for the voter model with RAU, synchronous and sequential asynchronous update dynamics in fully connected and various random uncorrelated networks (including scale-free networks). In all the cases explored we have found that $C(t)$ has an exponential tail with a characteristic time that scales as $\tau \sim N^{1/2}$ for random initial conditions.

\begin{figure}
 \includegraphics[height=0.47\textwidth,angle=-90,draft=false]{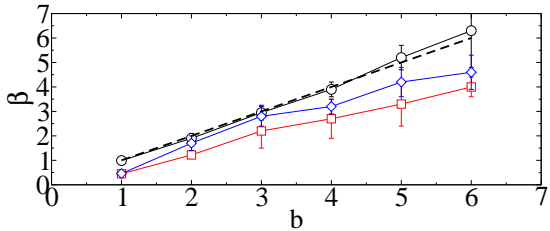}
 \caption{\label{rel_betas_topologies}Endogenous update. Relation of $\beta$, the exponent of the cumulative persistence distribution $C(t)\sim t^{-\beta}$, and $b$, the parameter in the function $p(\tau)=b/\tau$ for three different topologies; fully connected (circles), random with $\langle k\rangle=6$ (squares) and scale free  with $\langle k\rangle=6$ (diamonds) networks. As a guide to the eye we plot the curve $\beta=b$ with a dashed line.}
\end{figure}

We will now explore the consequences of the updates proposed above in the voter model: we examine the changes of the time evolution of $\langle \rho(t)\rangle$ and of the form of $C(t)$ with respect to the results with RAU dynamics.

{\em Fully connected networks.-} We first consider the case of a fully connected network with $p(\tau)=1/\tau$. Our results are summarized in Fig.~\ref{nocoev_update_rho_meanf}.
\begin{itemize}
\item[a)]\textit{Exogenous update}. The dynamics does not order the system: The average density of interfaces $\langle\rho(t)\rangle$ reaches a plateau in the thermodynamic limit. The density of interfaces averaged over surviving runs, $\langle\rho^*(t)\rangle$, reaches a plateau (inset of Fig.~\ref{nocoev_update_rho_meanf}), which is independent of the system size, showing that living runs stay, on average, on a dynamical disordered state.
\item[b)] \textit{Endogenous update}. The dynamics orders the system: The evolution of the average density of interfaces $\langle\rho(t)\rangle$ shows a power-law decay towards the absorbing configuration, with the same exponent for all system sizes. The average of the density over surviving runs, $\langle\rho^*(t)\rangle$ reaches a plateau, whose height decreases as $1/N$ indicating that in the thermodynamic limit it will be zero (Fig.~\ref{nocoev_update_rho_meanf}).
\end{itemize}
For the voter model with exogenous update the timescales are much larger than in the voter model with RAU, but it has the same qualitative behavior: the system doesn't order in the thermodynamic limit, but stays in a disordered dynamical configuration with asymptotic coexistence of both states. This contrasts with what happens with the endogenous update, where the timescales are also perturbed, but with the difference that a coarsening process occurs, slowly ordering the system. We have checked that the ensemble average of the magnetization $\langle m(t) \rangle=\frac{1}{N} \sum_{i=1}^N \langle s_i(t)\rangle$ is conserved for the exogenous update, whereas for the endogenous update this conservation law breaks down, as previously discussed in Ref.~\cite{tessone}. The non-conservation of the magnetization leads to an ordering process. The conservation law is broken due to the different average values of the persistence time in both populations of agents (+1 and -1) leading to different activation probabilities (agents changing to state $+1$ have larger $\langle \tau \rangle$ than the ones changing to state -1).

\begin{figure}
  \centering
  \includegraphics[height=0.22\textwidth,angle=-90]{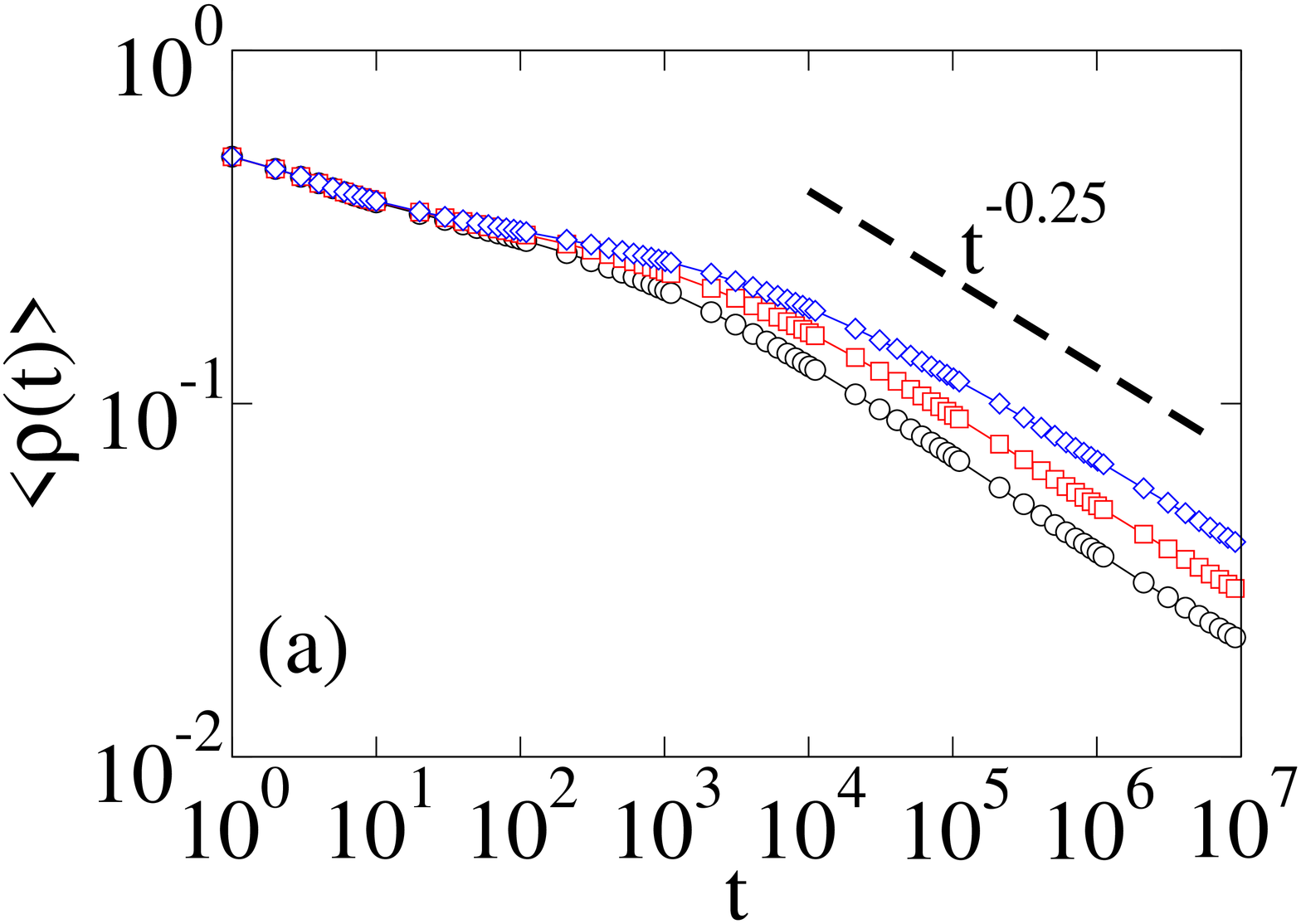}
  \includegraphics[height=0.22\textwidth,angle=-90]{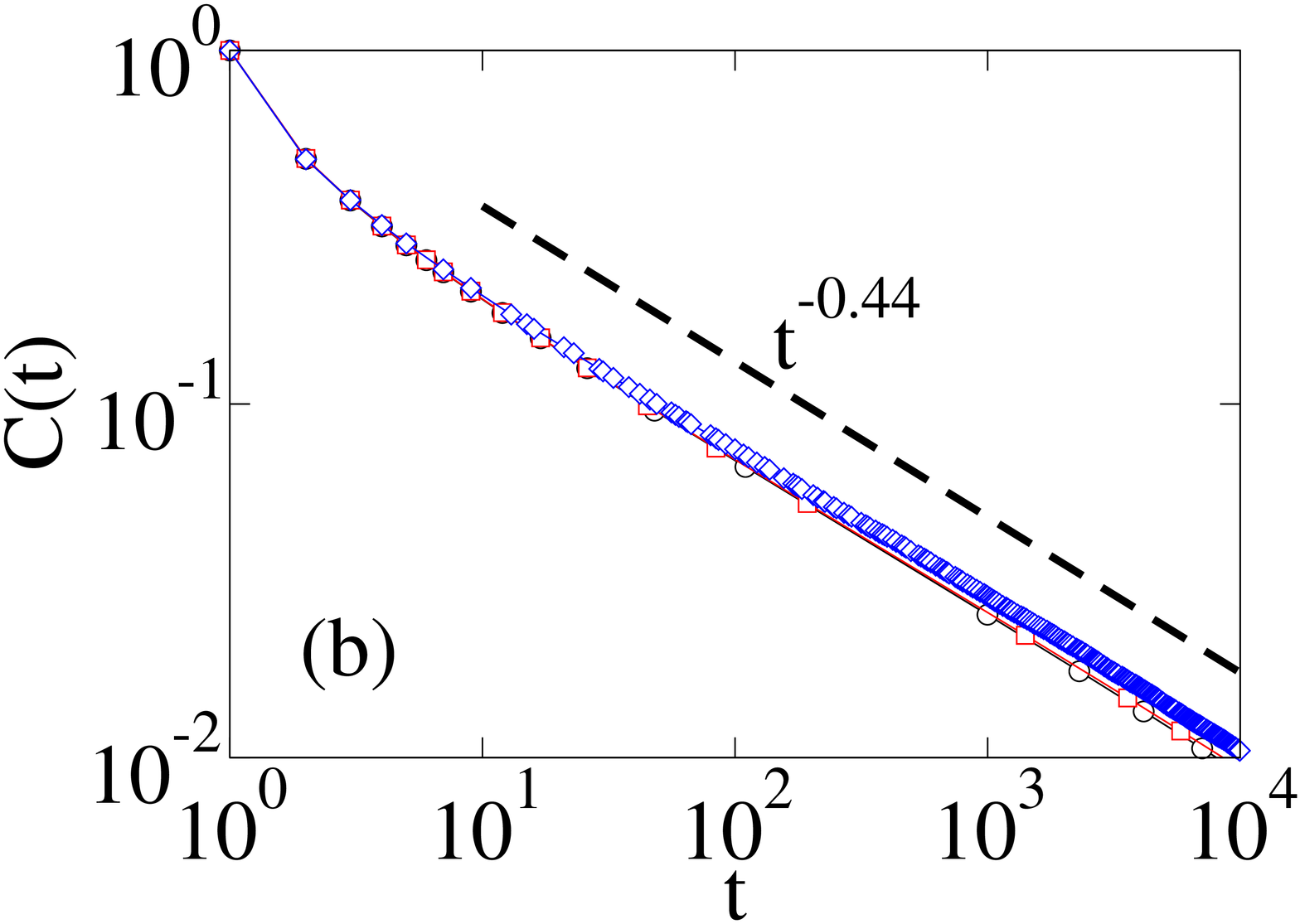}\\
  \includegraphics[height=0.22\textwidth,angle=-90]{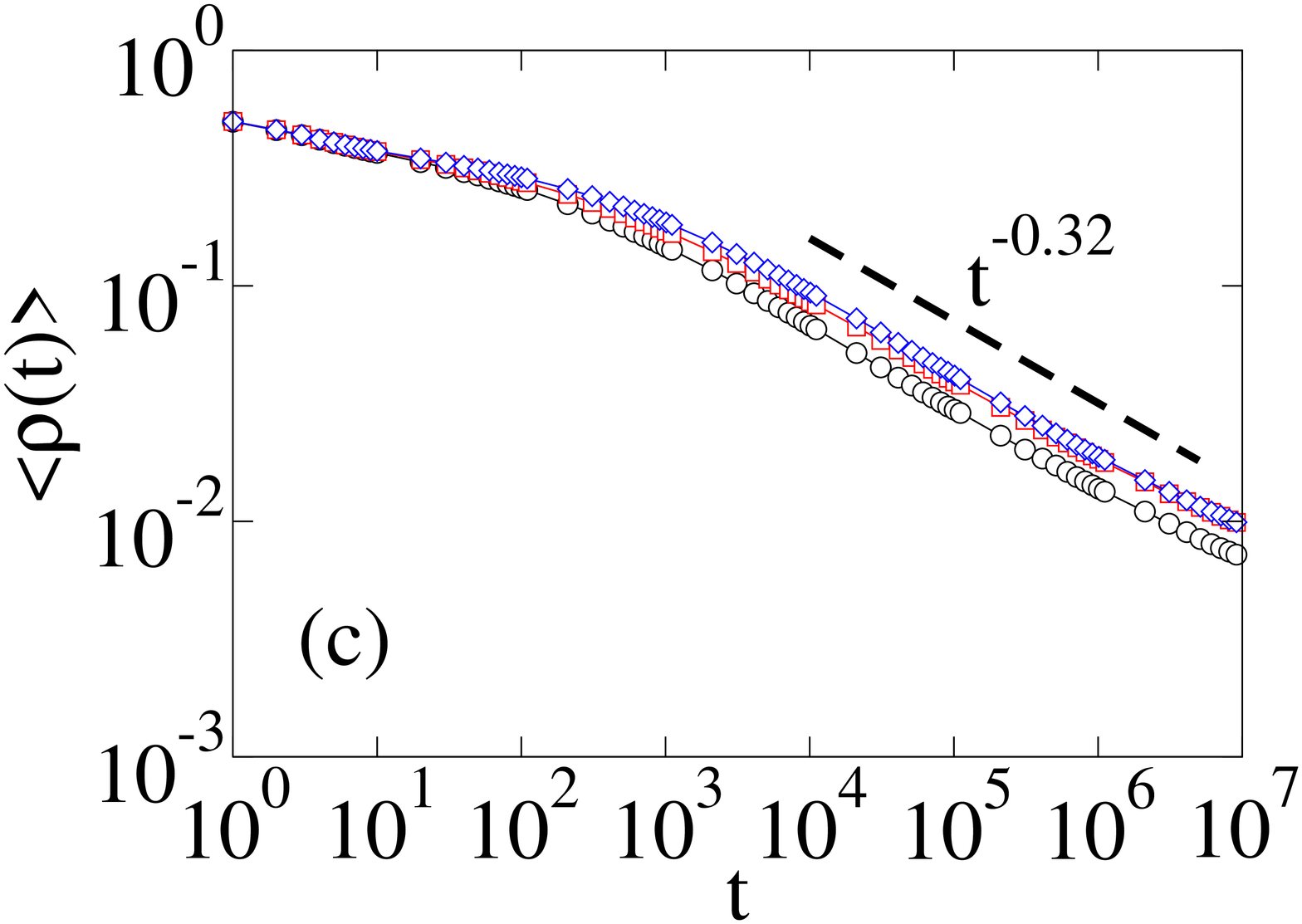}
  \includegraphics[height=0.22\textwidth,angle=-90]{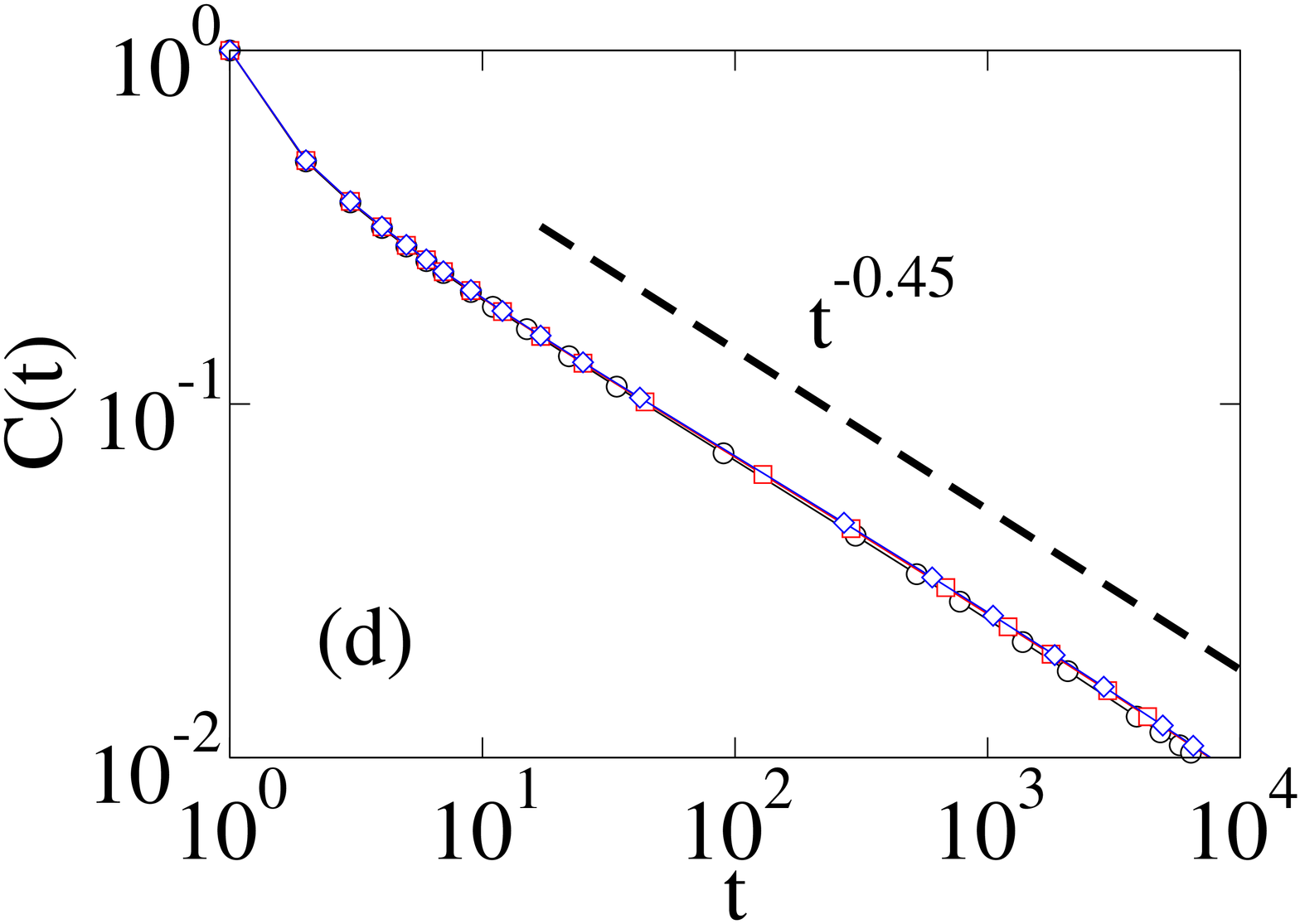}
  \caption{Endogenous update in complex networks: (a) average density of interfaces $\langle\rho(t)\rangle$, (b) cumulative persistence distribution $C(\tau)$ for an ER random network with $\langle k\rangle=6$; (c) average density of interfaces $\langle\rho(t)\rangle$, (d) cumulative persistence distribution $C(\tau)$ for a BA scale free network with $\langle k\rangle=6$. System sizes are 1000 (circles), 2000 (squares) and 4000 (diamonds); averages over $1000$ realizations. }
  \label{results_mem_updates_meanfield}
\end{figure}

Our results for the cumulative persistence distribution on a fully connected network with $p(\tau)=1/\tau$ are shown in Fig.~\ref{cdetau_meanf_coup_uncoup}. The distributions $C(t)$ show heavy tails consistent with a power-law of exponent $-1$. We expect the persistence $M(t)$ to be related to the activation probability $p(\tau)$. Neglecting the actual dynamics and assuming that at each update event, the agent changes state we can find an approximate relation between $M(t)$ and $p(\tau)$. Recall that $M(t)$ is the probability that an agent changes state (updating and changing state coincide in this approximation) $t$ timesteps after her last change of state. Therefore the probability that an agent has not changed state in $t-1$ timesteps is $1-\sum_{j=1}^{t-1}M(j)$ and the probability of changing state having persistence time $t$ is $p(t)$. Therefore we can write: $\left(1-\sum_{j=1}^{t-1}M(j)\right)p(t)=M(t)$, with $p(1)=M(1)$. Taking the continous limit and expressing this equation in terms of the cumulative persistence distribution we obtain $d\ln(C(t))=-p(t)dt$. Setting $p(\tau)=b/\tau$ the cumulative persistence distribution decays as a power law $C(t)\sim t^{-\beta}$ with $\beta=b$. Numerical simulations show that this approximation holds for a fully connected network for exogenous and endogenous updates (see Fig.~\ref{rel_betas_topologies}, exogenous update not shown).

\begin{table}
\begin{tabular}{c|c|c|}
  & $\langle\rho(t)\rangle\sim t^{-\gamma}$ & $C(t)\sim t^{-\beta}$ \\
\hline
 FC & $\gamma=0.99(1)$ &  $\beta=0.99(3)$ \\
 ER $\langle k \rangle=20$ & $\gamma=0.99(1)$ & $\beta=0.97(4)$ \\
 ER $\langle k \rangle=6$ & $\gamma=0.25(1)$ & $\beta=0.45(1)$ \\
 BA $\langle k \rangle=6$ & $\gamma=0.32(1)$ & $\beta=0.46(1)$ \\
\hline
\end{tabular}
\caption{Characteristic exponents of the average interface density $\langle\rho(t)\rangle$ and cumulative persistence distribution $C(\tau)$ for the voter model with endogenous update and activation probability $p(\tau)=1/\tau$. FC: fully connected network, ER: Erd\"os-R\'enyi random network, BA: Barab\'asi-Albert scale-free network.}
\label{table_exponents}
\end{table}

{\em Complex networks.-} To check the generality of the results obtained for a fully connected network we have performed simulations of the voter model with the exogenous and endogenous updates on complex networks such as Erd\"os-R\'enyi (ER) random networks \cite{Erdos59} and Barab\'asi-Albert (BA) scale-free networks \cite{Barabasi99}. We confirm the same qualitative results as in the fully connected network case in terms of ordering.

For the endogenous update the average interface density decays as a power law $\langle\rho(t)\rangle \sim t^{-\gamma}$ with an exponent $\gamma$ that depends on the interaction network, but not on the system size (see Figs.~\ref{results_mem_updates_meanfield}(a),(c)). Table~\ref{table_exponents} whows that the decay of $\langle \rho(t)\rangle$ is slower on complex networks, being slower in BA scale-free than in ER random networks. For random networks of high degree the behaviour of the model tends to the observed behaviour on a fully connected network, as expected. The density of interfaces averaged only over surviving runs $\langle \rho^*(t)\rangle$ displays a plateau for large times, whose height is inversely proportional to the system size (not shown). Therefore the system is coarsening and order is asymptotically reached in the thermodynamic limit, contrary to what happens with the standard updates in these networks \cite{vazquez_eguiluz}. For the exogenous update (data not shown) the average density of interfaces $\langle\rho(t)\rangle$ decays very slowly and slower for bigger system sizes and, in the thermodynamic limit stays on a dynamical disordered state, similarly to the fully connected network case.

The endogenous update gives rise to heavy tails in the persistence (see Figs.~\ref{results_mem_updates_meanfield}(b),(d) for the case $b=1$). In the case of complex topologies although the exponents of the power law tails of $C(t)$ are not given by $\beta=b$ as in the mean field case, it seems that they are proportional $\beta \propto b$ (Fig.~\ref{rel_betas_topologies}). The exponents for the case $b=1$ in different topologies are also summarized in Table~\ref{table_exponents} finding that they are smaller than 1 which is a signature of slow ordering, suggesting that the mean time to order is not well defined.

{\em Summary and Conclusions.-} Recent research on human dynamics has revealed the ``small but slow" paradigm \cite{small_but_slow,vazquez_non_poiss}, that is, the spreading of an infection can be slow despite the underlying small-world property of the underlying network of interaction. Here, with the help of a general updating algorithm for agent based models which can account for realistic interevent time distributions, we have shown that the competition of opinion can lead to slow ordering not only in small networks but also in the mean field case. By comparing the exogenous and the endogenous update we have pointed out the importance of a state dependent update. Our results provide a theoretical framework that bridges the empirical efforts devoted to uncover the properties of human dynamics with modeling efforts in opinion dynamics.

\begin{acknowledgments}
We acknowledge financial support from MICINN(Spain) through project FISICOS (FIS2007-60327). J.F.-G. aknowledges a predoctoral fellowship form the Government of the Balearic Islands.
\end{acknowledgments}


\end{document}